\documentclass{moriond}

\def\be{\begin{equation}}
\def\ee{\end{equation}}
\def\bea{\begin{eqnarray}}
\def\eea{\end{eqnarray}}

\newcommand{\Photo}{\includegraphics[height=35mm]{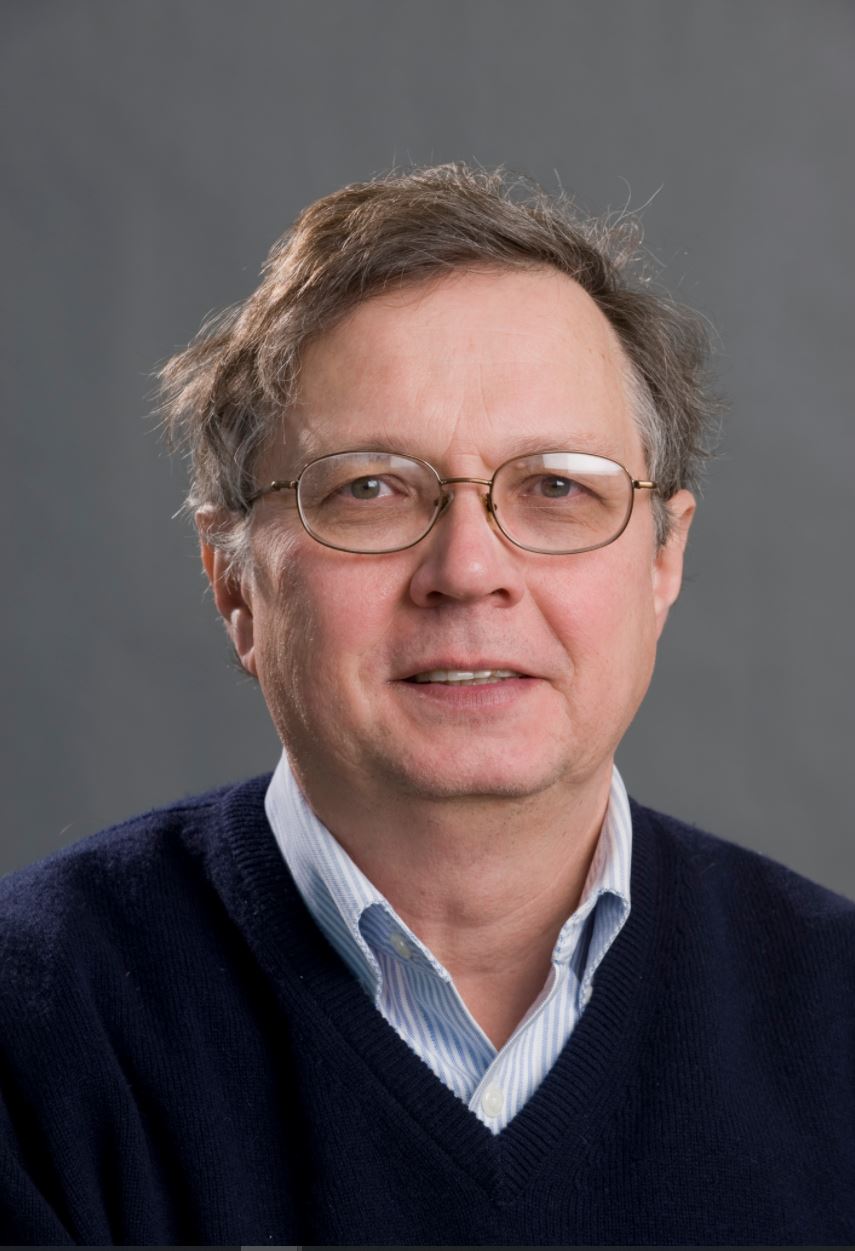}}

\begin{document}


  \begin{flushright}
  \makebox[1.9in]{FERMILAB-CONF-21-186-AD}

  \makebox[1.9in]{Moriond 2021 QCD}%

  \makebox[1.9in]{March 27 - April 3, 2021}
  \end{flushright}
  \vspace*{3.1cm}


\title{PRODUCTION OF EXOTIC HADRONS IN $p\overline{p}$ COLLISIONS 

AT 1.96 TEV AT D0}

\author{ PETER H. GARBINCIUS}

\address{on behalf of the D0 Collaboration \\ Fermi National Accelerator Laboratory \\
Batavia, Illinois 60510 USA}
\maketitle\abstracts{
D0 presents results on the hadronic production of $X(3872), \psi(2S), Z_c^\pm(3900),$ and the inclusive production of pentaquarks $P_c \rightarrow J/\psi \; p$.}

\section{Introduction}

\subsection{Exotic 4- and 5-quark particles}


In 1964, Gell-Mann and Zweig theorized that mesons were formed by quark-antiquark pairs and baryons by three quarks.  Both theories also allowed for mesons consisting of two quarks and two antiquarks and baryons consisting of four quarks plus an antiquark.  These exotic multi-quark states have been the subject of extensive theoretical and experimental studies~\cite{KRS,OSZ}. 
 
The first confirmed 4-quark state was the $X(3872)$ decaying into $J/\psi \: \pi^+ \pi^-$,  with quark content $c \overline{c} u \overline{u}$, discovered in 2001 by the Belle~\cite{Belle-X3872} experiment. 
The $X(3872)$ mass is within $<$ 0.25 MeV of the $D^0 \: \overline{D}^{*0}$ threshold, and since it has a large ($>$ 30 $\%$) branching fraction into $D^0 \: \overline{D}^{*0}$, there has been theoretical speculation that it is a weakly bound $D^0 \: \overline{D}^{*0}$ molecule.  However, such a molecule, with such a small binding energy, would have a large spatial extent $\approx$~10 fermi, and a large probability for dissociation by interactions with other particles produced at primary hadronic interaction vertices.  So, could the $X(3872)$ be composed of a $D^0 \: \overline{D}^{*0}$ molecule, or hadrocharmonium (a $c \overline{c}$ pair embedded in a cloud of $u \overline{u}$ light quarks), or a tightly bound tetraquark or diquark-antidiquark pair?

In 2015, the LHCb~\cite{LHCb-Pc} experiment discovered pentaquarks, with quark content $ u u d c \overline{c}$, in $\Lambda_b^0 \rightarrow K^- P_c^+$ with the pentaquark decay $P_c^+ \rightarrow J/\psi \: p$.

\subsection{The D0 Experiment and Analyses}




D0~\cite{D0-X3872,D0-Zc3900,D0-Pc} studied charmonium-like exotics decaying into $J/\psi \rightarrow \mu^+ \mu^-$ plus 1, 2, or 3 hadrons.  In $p \overline{p}$ collisions, $J/\psi$ are produced promptly, at the primary event vertex, either directly or in strong decays of higher-mass charmonium-like states, or non-promptly in $b$-hadron decays with the decay vertex displaced from the primary vertex, with proper decay length $c \tau \approx$ 0.46 $\mu$m.


The event sample was split into: ``displaced vertex" candidates from the decays of $b$-hadrons, based mainly on the significance of the transverse vertex separation $L_{xy}/\sigma_{L_{xy}}$, with the remaining events classified as ``primary vertex" candidates.

After correcting for the $b$-decay events feeding into the primary vertex candidates, we studied the prompt and non-prompt production of $X(3872)$ and $Z_c^\pm(3900)$ and compared with prompt and $b$-decay samples of charmonium $\psi(2S) \rightarrow J/\psi \; \pi^+ \pi^-$.

\section{\boldmath $X(3872)$ and Comparison with  \boldmath $\psi(2S)$ -- is the \boldmath $X(3872)$ a molecule?}

\begin{figure}
\begin{flushleft}
\includegraphics[width=6.5in]{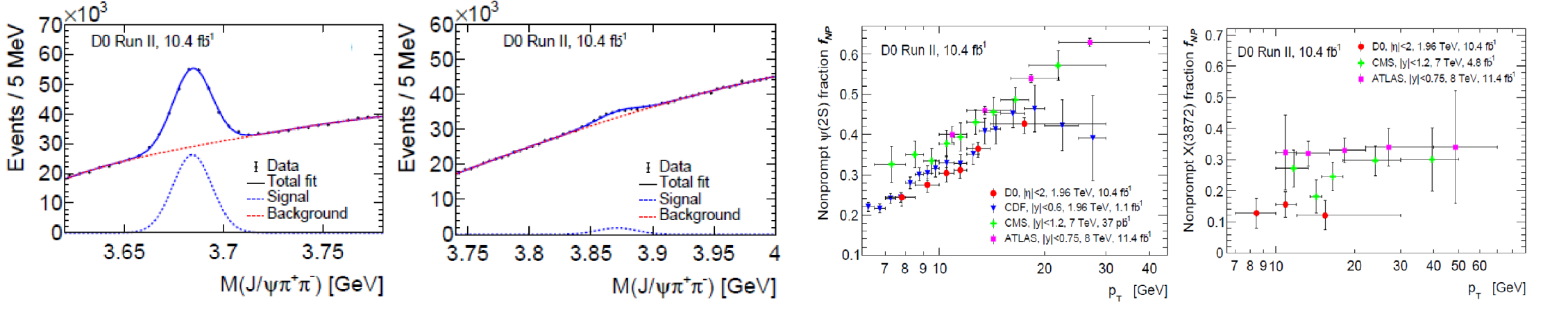}         
\caption{(left to right:) Mass spectra for (a) $\psi(2S)$ and (b) $X(3872)$, both decaying into $J/\psi \; \pi^+ \pi^-$;  
and $p_T$ distributions for the non-prompt fractions $f_{NP}$ for (c) $\psi(2S)$ and (d) $X(3872)$.}
\end{flushleft}
\end{figure}

We sorted the $X(3872)$ and $\psi(2S) \rightarrow J/\psi \; \pi^+ \pi^-$ candidates into bins in $p_T$ and 
pseudo-proper time,  $t_{pp} = \vec{L}_{xy} \cdot \vec{p}_T \; m/(p_T^2 \; c)$, and then fitted the for the $\psi(2S)$ and $X(3872)$ yields for that bin.  These yields vs. $t_{pp}$ were then fit for the non-prompt fractions $f_{NP}(\psi(2S))$ and $f_{NP}(X(3872))$. The prompt fractions are  $f_P = 1 - f_{NP}$.  Figure 1(c) shows the non-prompt fractions
$f_{NP}(\psi(2S),p_T)$, for ATLAS, CMS, CDF, and D0~\cite{D0-X3872}.  
In Fig. 1(d), $f_{NP}(X(3872),p_T)$ is constant with $p_T$, but $f_{NP}(LHC) \approx 3*f_{NP}(D0)$, The prompt production of $X(3872)$ may be suppressed at the LHC relative to the Tevatron due to the dissociation of the spatially large 
$D^0 \: \overline{D}^{*0}$ molecular component by interactions with the higher LHC hadron multiplicity at the primary vertex.  
This seems to be consistent with the recent LHCb $X(3872)$ production versus primary vertex multiplicity~\cite{LHCb-X3872}.

If $X(3872)$ were a molecule, it could be formed by the direct coalescence  of $D^0$ and $\overline{D}^{*0}$.  
Braaten {\sl et al.}~\cite{EB-1,EB-2} also considered production by the scattering 
$D^{*+} \: \overline{D}^{*0} \rightarrow \pi^+ \; D^0 \; \overline{D}^{*0}$ 
followed by the coalescence of $D^0$ $\overline{D}^{*0} \rightarrow X(3872)$.
They calculated that $f_\pi$ $\approx$ 14 $\%$ of the molecular component of the $X(3872)$, whether produced promptly in the primary hadronic vertex or in the decay of $b$-hadrons, would be accompanied by a $\pi^\pm$ with kinetic energy in the $X \pi$ center-of-mass 
reference frame, T($X\pi$) $<$ 11.8 MeV.  This estimate 
is strongly dependent on the binding energy of the $X(3872)$, which is still imprecisely known.  No such low-T enhancement is expected for the charmonium $\psi(2S) \; \pi$, so $\psi(2S) \; \pi$
is used to scale acceptances, efficiencies, and random coincidences for $X(3872) \; \pi$.  

Figure 2 displays the 5-track primary vertex
$\psi(2S) \: \pi$ (left) and $X(3872) \: \pi$ (center) with the T $<$ 11.8 MeV cut.  D0 found 44 $\pm$ 14 $\psi(2S) \: \pi$ events, consistent with the random coincidence rate.  After subtracting 6 expected random coincidences, 12 $\pm$ 16 $X(3872) \:\pi$ events are observed.  Applying the acceptances and the $f_\pi \approx 14 \%$ ratio, we expected between 245 and 730 events. Thus we have no evidence for the accompanying $\pi^\pm$ in the prompt production as expected for a molecular $X(3872)$ state.

The T($X\pi$) distribution for displaced 5-track vertex events is displayed in Fig. 2(right), showing 27 $\pm$ 12 events, including 2 random events for T($X \pi) < $ 11.8 MeV.  Again, scaling the total $X(3872)$ yield and applying the acceptances, 30-90 events would be expected.  
The displaced vertex $X(3872) \: \pi$
events, although consistent with the expected yield, are also within 2$\: \sigma$ of the null-hypothesis for this low-T($X\pi$) molecular scattering process.  Of course, the $X(3872)$ could have only a small molecular component and be consistent with these observations.

\begin{figure}
\begin{flushleft} 
\includegraphics[width=6.25in]{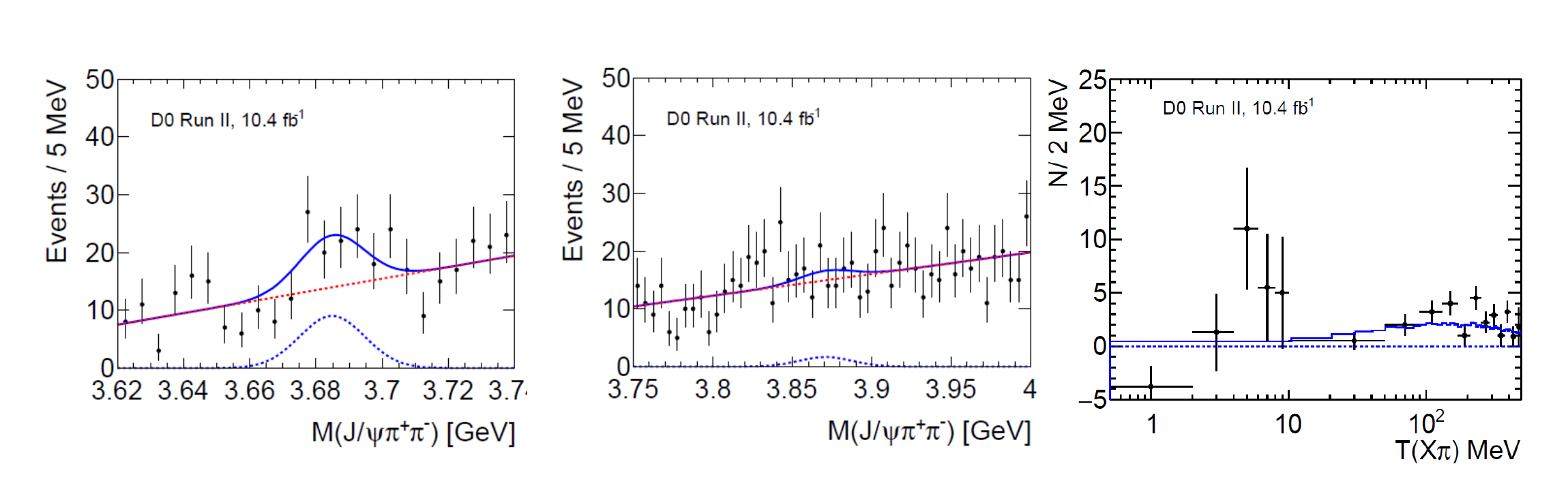}     
\caption{$J/\psi \; \pi^+ \pi^-$ mass distributions for $J/\psi \; \pi^+ \pi^- \pi^\pm$ with T $<$ 11.8 MeV cut for $\psi(2S)$ (left) and $X(3872)$ (center) for the primary vertex event selection.  (right) The T($X \pi$) distribution for the displaced vertex event selection for $X(3872) \pi$.}
\end{flushleft}
\end{figure}

\section{Production of \boldmath $Z_c^\pm(3900)$}

The $Z_c^+(3900)$  has a quark composition of $c \overline{c} u \overline{d}$ and decays into $J/\psi \; \pi^+$.  D0~\cite{D0-Zc3900} searched for 
$Z_c^\pm(3900)$ in 4-track vertex events with $Z_c^\pm \; \pi^\mp$ arising from both prompt and $b$-hadron decay processes.  We performed a coarse scan in six 100 MeV wide mass bins for the parents of the $Z_c(3900)$, e.g. the previously seen 
$\psi(4260) \rightarrow Z_c(3900) \; \pi$.

Figure 3(left) shows the M($J/\psi \; \pi^\pm$) spectrum for the 4.2 $<$ M($J/\psi \; \pi^\pm \pi^\mp$) $<$ 4.3 GeV bin for displaced vertices, the only bin with a significant $Z_c(3900)$ yield.  We found 376 $\pm$ 76 $Z_c(3900)$ events with a significance of $5.2 \: \sigma$. Figure 3 also shows the $Z_c(3900)$ yields for each of the 6 parent mass bins for the displaced and primary vertex events.

\begin{figure}
\centering 
\includegraphics[width=6.25in]{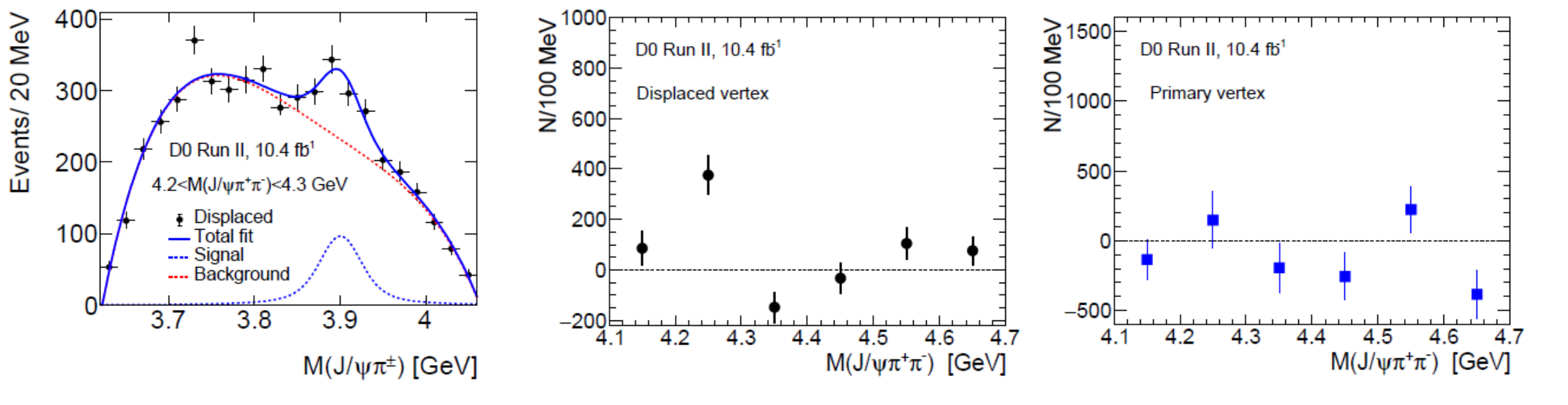}     
\caption{(left) The M($J/\psi \; \pi^\pm$) distribution for 4.2 $<$ M($J/\psi \; \pi^+ \; \pi^-$) $<$ 4.3 MeV for displaced vertex events, the only mass range with a $Z_c^\pm(3900)$ yield statistically greater than zero; the $Z_c^\pm(3900)$ yields for the six M($J/\psi \; \pi^+ \; 
\pi^-$) mass bins, (center) displaced vertex events, and (right) primary vertex events. }
\end{figure}

\section{Inclusive Pentaquarks}

Until now, only LHCb~\cite{LHCb-Pc} has observed pentaquarks, quoting masses, widths, and relative yields for three $P_c$ states at 4312, 4440, and 4450 MeV. 
D0~\cite{D0-Pc} searched for inclusive production of pentaquarks in $P_c \rightarrow J/\psi \; p$ without reconstructing a $b$-baryon parent.
The D0 mass resolution is insufficient to separate the $P_c(4440)$ and $P_c(4457)$ states.  We fixed the masses and widths at the LHCb quoted values and convoluted with the D0 mass resolution for our fits.  We assumed an incoherent sum of $P_c(4440)$ and 
$P_c(4457)$, but allowed their relative yields to vary.  Figure 4(left) shows the $J/\psi \; p$
 mass spectrum for displaced vertices, finding a total of 
N(4440) + N(4457) = 830 $\pm$ 206 events, with significance of $3.2 \: \sigma$ (stat. + syst.).  
The ratio of N(4440)/(N(4440)+N(4457)) = 0.61 $\pm$ 0.23, consistent with the LHCb ratio of 0.68.
D0 found only 151 $\pm$ 186 events for displaced vertex $P_c(4312)$.   


\begin{figure}
\centering  
\includegraphics[width=4.3in]{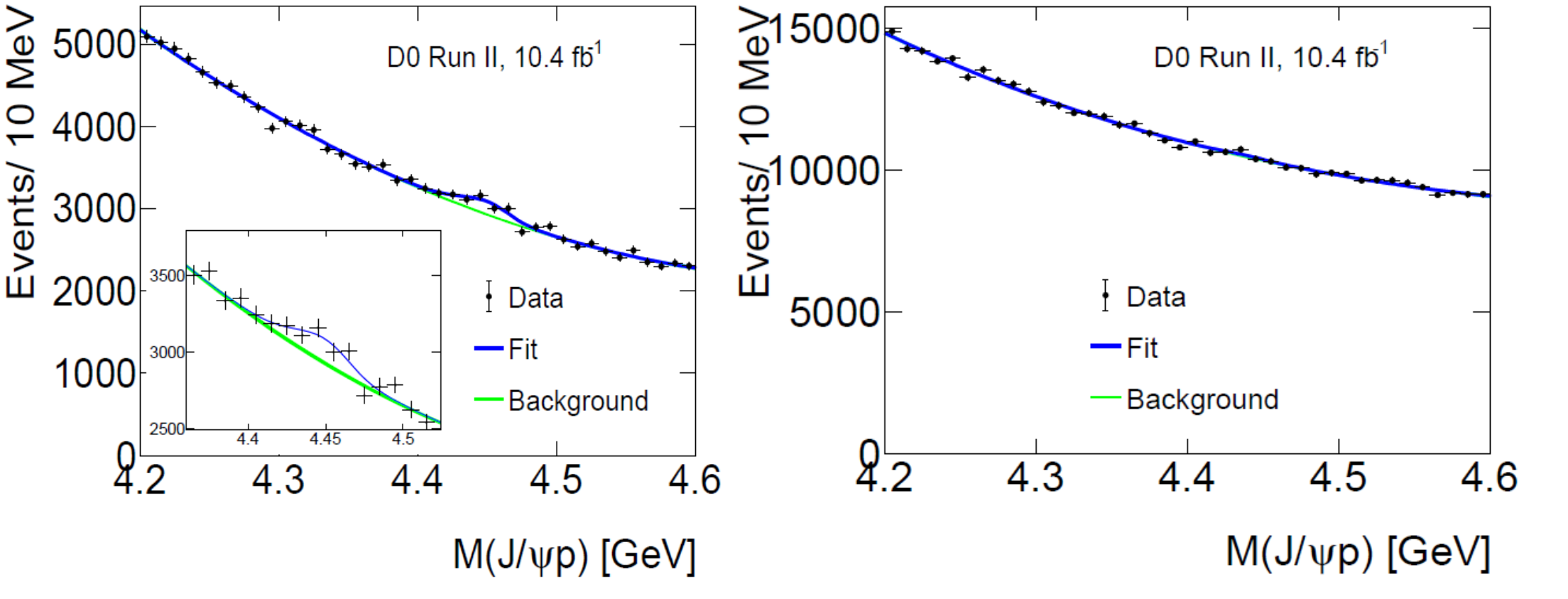}    
\caption{D0 Inclusive Invariant Mass($J/\psi \: p$) distributions:  (left:) showing an unresolved peak at the average mass  of the $P_c(4440)$ and $Pc(4457)$ states with significance of $3.2 \: \sigma$ (statistical + systematic) for displaced vertex events;  (right:)  no significant peak was observed for primary vertex events, nor for the $Pc(4312)$ state.}
\end{figure}

\section{Summary Highlights}

The non-prompt fraction $f_{NP}(X(3872))$ at the Tevatron is $\approx$ 3 times lower than at the LHC, indicating that prompt production of a 
$D^0  \: \overline{D}^{*0}$ molecular component of the $X(3872)$ may be more readily dissociated by interactions with the higher primary vertex multiplicity at the LHC.

We found no significant evidence for $\overline{D}^{*+} + D^{*0}$ to produce a co-moving $X(3872$ and $\pi^+$.  
 
We observed $Z_c^\pm$ production consistent with the sequential decay of $b$-hadrons $\rightarrow \psi(4260) \rightarrow Z_c(3900)^\pm  
\pi^\mp $, but found no significant evidence for $Z_c(3900)$ via decays of prompt $\psi(4260)$ or via decays of other parents of $Z_c(3900)^\pm  \pi^\mp$ within the 4.1-4.7 GeV mass range.

We observed an inclusive unresolved peak corresponding to the $P_c(4440) + P_c(4457)$ pentaquarks from the decay of $b$-baryons 
with significance = $3.2 \: \sigma$ (stat. + syst.).  This was the first confirmation of the LHCb pentaquark discovery.
We found no significant evidence for prompt $P_c(4414) + P_c(4457)$ or for the $P_c(4312)$ from either prompt or $b$-baryon decay processes.

\section*{Acknowledgments}
 
The D0 Collaboration thanks the staffs at Fermilab and collaborating institutions,
and acknowledges support of the many international funding agencies which made this experiment possible.  


\section*{References}   

\begin{thebibliography}{99}

\bibitem{KRS} M. Karliner, J.L. Rosner, and T. Skwarnicki, 
{\sl Ann. Rev. Nucl. Part. Sci.} {\bf 68}, 17 (2018).
     
\bibitem{OSZ} S.L. Olsen, T. Skwarnicki, and D. Zieminska, 
 {\sl Rev. Mod. Phys.} {\bf 90}, 01003 (2018).

\bibitem{Belle-X3872} S.-K. Choi {\sl et al.}, (Belle Collaboration), 
{\sl Phys. Rev. Lett.} {\bf 91}, 262001 (2003).

\bibitem{LHCb-Pc} R. Aaij {\sl et al.}, (LHCb Collaboration), 
{\sl Phys. Rev. Lett.} {\bf 122}, 222001 (2019).

\bibitem{D0-X3872} V.M. Abazov {\sl et al.}, (D0 Collaboration),
{\sl Phys. Rev.} D {\bf 102}, 072005 (2020).

\bibitem{D0-Zc3900} V.M. Abazov {\sl et al.}, (D0 Collaboration),         
{\sl Phys. Rev.} D {\bf 100}, 012005 (2019).


\bibitem{D0-Pc} V.M. Abazov {\sl et al.}, (D0 Collaboration),
arXiv:1910.11767, unpublished. This $P_c$ analysis was not accepted for publication, the main reason being the detailed set of cuts applied.  We modified the cuts slightly from earlier exotic meson analyses due to a new, more relaxed event pre-selection.

\bibitem{LHCb-X3872} R. Aaij {\sl et al.}, (LHCb Collaboration),  
{\sl Phys. Rev. Lett.} {\bf 126}, 092001 (2021). 

\bibitem{EB-1} E. Braaten, L.-P. He, and K. Ingles, 
{\sl Phys. Rev.} D {\bf 100}, 094006 (2019).
 
\bibitem{EB-2} E. Braaten, L.-P. He, and K. Ingles, 
{\sl Phys. Rev.} D {\bf 100}, 074028 (2019). 


\end{thebibliography}
\end{document}